\documentstyle[twoside,fleqn,espcrc2,epsf]{article}
\newcommand{\be}{\begin{equation}}
\newcommand{\ee}{\end{equation}}
\newcommand{\bea}{\begin{eqnarray}}
\newcommand{\eea}{\end{eqnarray}}
\newcommand{\bean}{\begin{eqnarray*}}
\newcommand{\eean}{\end{eqnarray*}}

\newcommand{\AmS}{{\protect\the\textfont2
  A\kern-.1667em\lower.5ex\hbox{M}\kern-.125emS}}

\title{Heavy Quarkonia from Anisotropic and Isotropic Lattices
\thanks{talk presented by T.~Manke}}

\author{
{CP-PACS Collaboration}:
A.~Ali~Khan\address{Center for Computational Physics,
University of Tsukuba, Tsukuba, Ibaraki 305-8577, Japan}, 
S.~Aoki\address{Institute of Physics,
University of Tsukuba, Tsukuba, Ibaraki 305-8571, Japan},
R.~Burkhalter$^{\rm a,b}$, 
S.~Ejiri$^{\rm a}$, 
M.~Fukugita\address{Institute for Cosmic Ray Research, 
University of Tokyo, Tanashi, Tokyo 188-8502, Japan}, 
S.~Hashimoto\address{High Energy Accelerator Research Organization (KEK), 
Tsukuba, Ibaraki 305-0801, Japan}, 
N.~Ishizuka$^{\rm a,b}$,
Y.~Iwasaki$^{\rm a,b}$, 
K.~Kanaya$^{\rm a,b}$, 
T.~Kaneko$^{\rm a}$, 
Y.~Kuramashi$^{\rm d}$,
T.~Manke$^{\rm a}$,
K.~Nagai$^{\rm a}$,
M.~Okawa$^{\rm d}$, 
H.P.~Shanahan\address{DAMTP, 21 Silver St., University of Cambridge,
Cambridge, CB3 9EW, England, U.K.},
A.~Ukawa$^{\rm a,b}$ and T.~Yoshi\'e$^{\rm a,b}$ }
       
\begin{document}

\begin{abstract}
We report on recent results for the spectrum of heavy quarkonia. 
Using coarse and anisotropic lattices we achieved an unprecedented 
control over statistical and systematic errors for higher excited 
states such as exotic hybrid states.
In a parallel study on isotropic lattices we also investigate
the effect of two dynamical flavours on the spin structure 
of charmonium and bottomonium for several symmetric lattices.

\end{abstract}
%
\maketitle
\section{INTRODUCTION}
Heavy quarkonia play an important role for the theoretical understanding of
QCD. Their non-relativistic character has frequently been used to perform 
efficient lattice simulations and has triggered many detailed studies of 
systematic errors such as lattice spacing artefacts, relativistic corrections 
and quenching effects. As an additional advantage there exists a wealth of
experimental data, which provide an ultimate check on different improvement
programmes. 
Moreover, lattice calculations have also resulted in predictions from first
principles for heavy $Q\bar Qg$-states containing an additional gluonic 
excitation \cite{cmhybrid,ukqcd_hybrid97,milc}.
However, those attempts were hampered by the rapidly decaying signal-to-noise ratio of such high-energetic
hybrid states on conventional lattices.
 
More recently anisotropic lattices have been used to circumvent this problem
by giving the lattice a fine temporal resolution whilst maintaining
a coarse discretisation in the spatial direction
\cite{morning_glue98,kuti_hybrid98}. 
In a previous study we already reported on first NRQCD results 
for charmonium and bottomonium hybrid states from anisotropic lattices
\cite{cppacs_hybrid98}. In Section 2 we report on further applications of
those methods and study also other excitations in heavy quarkonia more
carefully. 
 
In quenched simulations without dynamical sea quarks the strong 
coupling does not run as in nature and therefore one cannot reproduce
experimental quantities at all scales.
Observed deviations of the quenched hadron spectrum from experiment have been
reported previously and an improvement has been noticed
once dynamical quarks are inserted into the gluon background 
\cite{cppacs98}. Here we study unquenching effects for heavy quarkonia
and report on our results from isotropic lattices in Section 3.

\section{EXCITED QUARKONIA}
To study excited states with small statistical errors it is mandatory
to have a fine resolution in the temporal lattice direction, along
which we measure the multi-exponential decay of meson correlators.
To this end we employ an anisotropic and spatially coarse gluon action:

\begin{eqnarray}
S = - \beta \sum_{x, {\rm i > j}} \xi^{-1}\left\{\frac{5}{3}P_{\rm i j} -
\frac{1}{12}\left(R_{\rm i j} + R_{\rm j i}\right)\right\} - \nonumber \\
 \beta \sum_{x,{\rm i}} \xi \left\{\frac{4}{3}P_{\rm i t} -
\frac{1}{12}R_{\rm i t}\right\}~~.
\label{eq:glueaction}
\end{eqnarray}
Here $(\beta,\xi)$ are two parameters, which determine the gauge coupling
and the anisotropy of the lattice. Action (\ref{eq:glueaction}) 
is Symanzik-improved and involves plaquette terms, $P_{\mu\nu}$, as well as
rectangles, $R_{\mu\nu}$. It is designed to be accurate up to ${\cal
O}(a_s^4,a_t^2)$, classically. 
To reduce the radiative corrections we used self-consistent mean-field
improvement for both spatial and temporal links. With this prescription
we expect only small deviations of $\xi$ from its tree-level value $a_s/a_t$.

\begin{figure}[t]
\hbox{\epsfxsize = 90mm \epsfysize = 66mm \hskip -10mm \epsffile{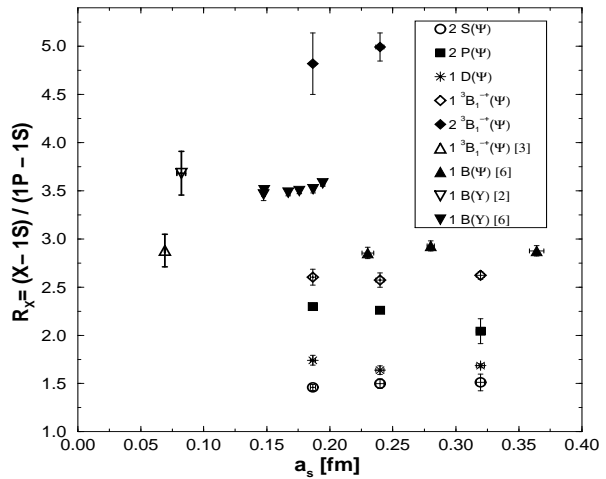}}
\vskip -7mm
\caption{Scaling analysis for excited Quarkonia. We plot the ratio $R_X$
against the spatial lattice spa\-cing for different states X = 2S,2P,1D
and magnetic hybrids (1B,2B).}
\label{fig:RX}
\vskip -7mm
\end{figure}

For the heavy quark propagation in the gluon background 
we used the NRQCD approach on anisotropic lattices as described in 
\cite{ron_lat98}. 
From the quark propagators we construct meson correlators for bound
states with quantum numbers $S(0,1) \times  L(0,1,2)$ and for 
hybrid states with additional gluonic excitation.
For example, the spin-singlet operators read
\be
\bar Q^\dag Q~,~
\bar Q^\dag \Delta_i Q~,~ 
\bar Q^\dag \Delta_j \Delta_k Q \mbox{~~and~~}
\bar Q^\dag B_i Q~.  
\ee
%
Within the NRQCD approach it is paramount to establish a scaling region
at finite lattice spacing. Our results in Fig.~\ref{fig:RX}
demonstrate that we succeeded in finding such a window.
As we only measure excitation energies relative to the
ground state it is natural to present our results as the ratio
$R_X=(X-1S)/(1P-1S)$, which gives the normalized splitting of state X
above the 1S. 

For the lowest lying hybrid excitations, $c\bar cg$ and $b\bar
bg$,  our results from leading order NRQCD \cite{cppacs_hybrid98} are in
excellent agreement with previous calculations on isotropic lattices 
\cite{ukqcd_hybrid97,milc}, but with much smaller errors. This is the combined
success of anisotropic and coarse lattices with a clear signal over many
time\-slices at small computational cost.
Here we have also checked the spin-averaged hybrid against possible 
finite volume errors, temporal lattice spacing artefacts 
and relativistic corrections, but we did not find any significant effect.

In Fig. \ref{fig:RX} we have also shown our new results for
higher radial excitations and D-states ($L=2$). 
Since all the spin corrections up to ${\cal O}(mv^6)$ are now included in our analysis we can
also determine the spin-structure very accurately. In particular we were able
to extract the exotic hybrid quarkonia, $^3B_1^{-+}$, explicitly for the first
time from NRQCD.  
Our data indicates that the spin splittings in hybrid states 
are enlarged compared to P-states, whereas the D-state splittings
are much suppressed. A more detailed discussion of the spin structure in 
quarkonia is presented elsewhere~\cite{qcd99}. 

The dominating systematic error for all the predictions in this section 
is an uncertainty in the scale as a result of the quenched approximation.
This is not yet controlled and we find a variation of 10-20\%,
depending on which experimental quantity is used to set the scale.

\section{SPIN STRUCTURE AND SEA-\\ QUARK EFFECTS}

To study sea quark effects in heavy quarkonia we employ an ensemble 
of isotropic lattices, where the gauge configurations were generated 
from an RG improved gluon action and tadpole-improved SW fermions for two
flavours of light sea-quarks \cite{cppacs98}. 
We then performed an NRQCD calculation of S- and P-states at two different
gauge couplings of $\beta=1.80$ and $\beta=1.95$, corresponding to 
$a\approx 0.2$ fm and $a\approx 0.15$ fm. 
In addition to the chiral extrapolation from four different sea quark masses,
we also compared our results directly to quenched simulations at 
the same lattice spacing. 

\begin{figure}[t]
\hbox{\epsfxsize = 90mm  \epsfysize = 66mm \hskip -10mm \epsffile{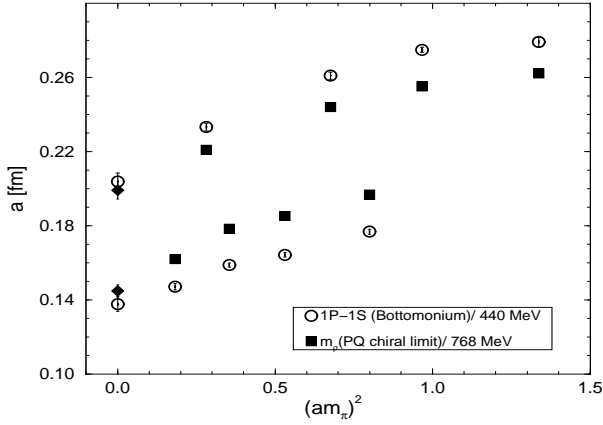}}
\vskip -10mm
\caption{Chiral extrapolation of the lattice spacing at two different
couplings determined from $m_\rho$ (squares) and $1P-1S$ in Bottomonium
(circles).}
\vskip -7mm
\label{fig:a_vs_ampi2}
\end{figure}
In Fig. \ref{fig:a_vs_ampi2} we see an encouraging trend
for the lattice spacings from different physical quantities
to agree much better in the chiral limit. This is an improvement
over quenched simulations, where one has large uncertainties in the scale.

Furthermore, we observed a clear shift upwards of the hyperfine splittings
as the sea quark mass is decreased.
\begin{figure}[t]
\hbox{\epsfxsize = 90mm  \epsfysize = 66mm  \hskip -10mm \epsffile{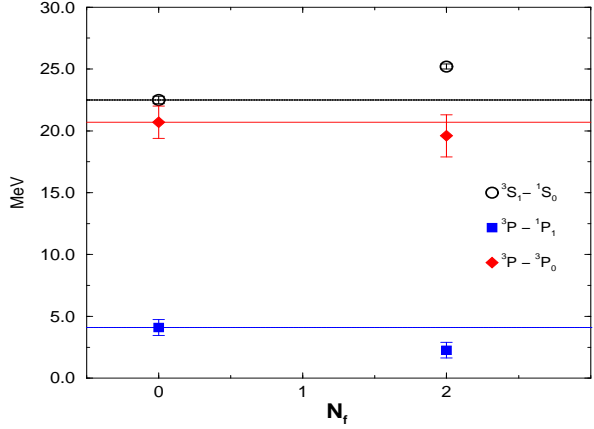}}
\vskip -10mm
\caption{Spin structure of Bottomonium. Here we plot results from quenched and 
two-flavour QCD at the same lattice spacing $a\approx 0.164$ fm.}   
\label{fig:unquenching}
\vskip -7mm
\end{figure}
For the Bottomonium we find an effect of about 3 MeV in a direct comparison to
a quenched calculation using an identical formulation of tadpole-improved
NRQCD with accuracy ${\cal O}(mv^6)$. This is a 10-$\sigma$ effect (or 10\%)
as shown in  Fig. \ref{fig:unquenching}.  In Charmonium the hyperfine splitting
is also raised by about $+15\%$, i.e. to around 60 MeV in the chiral limit on
our coarsest lattice. 

In P-states one expects a different situation, since their wavefunctions
va\-nish at the origin. 
Indeed, the very small hyperfine splitting ($^3P-{^1}P_1$) is further
suppressed ($\approx$ 3-$\sigma$) on dynamical configurations and there is
no resolvable shift for the fine structure, e.g. $^3P-{^3}P_0$. This validates the
quenched approximation for such quantities. 

For the Bottomonium system we are 
presently performing a similar analysis at $\beta=2.1$ ($a\approx 0.1$ fm)
to determine whether those observations still hold on finer lattices.

\vskip 20pt

This work is supported in part by Grants-in-Aid
of the Ministry of Education,
Science and Culture (Nos.~09304029, 10640246, 10640248, 10740107,
11640250, 11640294, 11740162).
SE and KN are JSPS Research Fellows.
AAK, HPS and TM are supported by the Research for the Future
Program of JSPS, and HPS also by the Leverhulme foundation.

\end{document}